\documentclass[pra, 10pt, twocolumn,letter, floatfix]{revtex4}

\usepackage{amsmath}
\usepackage{amsfonts}
\usepackage{amssymb}
\usepackage[T1]{fontenc}       
\usepackage{graphicx}
\usepackage{dcolumn}
\usepackage{bm}
\begin{document}

\title[Equation of motion JAGP] {Equation of motion theory for excited states in variational Monte Carlo and the Jastrow antisymmetric geminal power in Hilbert space}

\author{Luning Zhao$^{1,2}$}
\author{Eric Neuscamman$^{1,2,}$\footnote[1]{Electronic mail: eneuscamman@berkeley.edu}}
\affiliation{$^1$Department of Chemistry, University of California, Berkeley, California 94720, USA\\
             $^2$Chemical Sciences Division, Lawrence Berkeley National Laboratory, Berkeley, California 94720, USA}

\date{\today}


\begin{abstract}
An equation of motion formalism for excited states in variational Monte Carlo is derived and
a pilot implementation for the Jastrow-modified antisymmetric geminal power is tested.
In single excitations across a range of small molecules,
this combination is shown to be intermediate in accuracy between configuration interaction singles
and equation of motion coupled cluster with singles and doubles.
For double excitations, energy errors are found to be similar to those for coupled cluster.
\end{abstract}

\maketitle

\section{Introduction}
\label{sec:introduction}

Molecular spectroscopy, especially laser spectroscopy, has revolutionized research in physical chemistry.
Using laser spectroscopy, chemists can measure the spectra and photochemical dynamics of molecules,
identifying transition states and studying the rate of light-initiated chemical reactions with high spectral
and time resolution.
However,
the correct interpretation of
these spectra often
requires
detailed
knowledge of electronic excited states, such as vertical excitation
energies,
spin multiplicity and state symmetry.
Furthermore, dark states, which
can be hard to access experimentally due to their shared symmetry with
the ground state, play very important roles in quantum dynamics\cite{Blatt:2013:quantum_dynamics}.
These difficulties provide a major opportunity for theoretical predictions to contribute to our
understanding of quantum chemistry.

Unfortunately, theoretical methods in quantum chemistry are often less effective for excited states than they
are for the ground state, in no small part because they lack an equivalently robust and established variational
principle with which their wave function approximations may be optimized.
In many excited states, this difficulty is compounded by the breakdown of the independent particle approximation,
as occurs for example in states with significant doubly excited character.
While states with spatial symmetry or spin multiplicity differing from the ground state can be treated variationally
in a relatively straightforward manner, states in molecules lacking symmetry or that share a representation with
the ground state require a more general approach.



One such approach exists in the class of methods based on configuration interaction (CI), including
CI itself \cite{Kutselnigg:1988:ci} as well as the 
complete active space self-consistent field (CASSCF) method \cite{MolElecStruc} and its CI-based (MRCI) \cite{Butsher:1978:mrci}
and perturbative (CASPT2) extensions \cite{ModernElecStruc}.
CASSCF-based methods are among the most robust available and have the advantage of systematic convergence via expansions of the active space,
but they suffer from the need for state-averaging and combinatorially growing costs.
More recent methods offering systematic convergence include full configuration interaction quantum Monte Carlo (FCI-QMC)
\cite{Blunt:2015:krylov_fciqmc,Blunt:2015:excited_state_fciqmc}
and the density matrix renormalization group \cite{Chan:2008:dmrg_beta_carotene,Chan:2015:excited_state_geometry_optimization},
although these also have combinatorially growing costs in general.

Another approach to excited states is to apply linear-response or equation-of-motion theory,
which build excited state approximations using linear combinations of an approximate ground state ansatz's first derivatives.
Such methods include
configuration interaction singles (CIS) \cite{Martin:2005:LR},
time-dependent density functional theory (TDDFT) \cite{Martin:2005:LR}, 
equation-of-motion coupled cluster with singles and doubles (EOM-CCSD) \cite{Krylov:2008:eom_ccsd},
and linear-response DMRG \cite{Chan:2014:LR_DMRG}. 
These methods typically have more favorable cost-scalings than those based on CASSCF, allowing them to reach larger molecules,
but they usually lack the advantage of systematic improvability.


We recently introduced an excited state variational principle \cite{Zhao:2016:direct_tar} as a complementary alternative to these paradigms
and used it to explore the effectiveness of optimizing a Hilbert-space Jastrow antisymmetric geminal power (JAGP) wave function
\cite{Neuscamman:2012:sc_jagp,Neuscamman:2013:hilbert_jagp} for individual excited states.
The ability of this approach to rival, and for double excitations surpass, the accuracy of EOM-CCSD raised a natural question:
is the JAGP only accurate under direct targeting, or would this accuracy be maintained if used in a more traditional linear response framework?
In the present study we provide answers to this question by developing an equation-of-motion (EOM) formalism for variational Monte Carlo (VMC)
that is capable of working with the JAGP.



While we limit our current investigation to the JAGP ansatz, we stress that the EOM-VMC formalism itself is amenable to use with any wave function
ansatz to which the VMC linear method (LM) \cite{Nightingale:2001:linear_method,Umrigar:2005:lm,UmrTouFilSorHen-PRL-07,TouUmr-JCP-07,TouUmr-JCP-08}
optimization is applicable.
We thus seek to add the EOM approach to the growing toolkit of quantum Monte Carlo (QMC) methods for excited states, which has recently
been expanded by an excited state variational principle \cite{Zhao:2016:direct_tar},
state-averaging approaches \cite{Fillippi:2009:sa_qmc,Filippi:2013:excited_state_geometry_optimization}, 
and approaches based on FCI-QMC \cite{Blunt:2015:krylov_fciqmc,Blunt:2015:excited_state_fciqmc}.

As the data we present will show, EOM-JAGP provides a performance in relative excitation energies (that is to say differences between excitation energies)
that is intermediate between that of CIS and EOM-CCSD, a finding that is perhaps not surprising given that the complexity of the ground state ansatz
and therefore the size of its linear response space is also intermediate between these methods.
Due to the relatively high quality of the JAGP for ground states, however, which is much closer to that of CCSD than to restricted Hartree Fock (RHF),
the absolute excitation energies from EOM-JAGP are typically overestimated, even more so than is common in other EOM theories such as CIS.
While we make a distinction in this paper between a method's relative accuracy, which is not affected by EOM's typical ground-state bias, and absolute
accuracy, it is clear that reducing ground state bias in EOM-JAGP and EOM methods more generally is an important priority for future research.

This paper is organized as follows.
We begin with a review of the equation-of-motion
formalism as derived from time dependent perturbation theory (Section \ref{sec::eom}).
We then explain the use of LM technology
to evaluate the EOM Hamiltonian and overlap matrices and discuss why a naive approach to Monte Carlo sampling can
be problematic in this evaluation (Section \ref{sec::theory_HS}).
Having laid out the general formalism, we introduce the JAGP ansatz and discuss the nature of its linear response space
(Sections \ref{sec::jagp} and \ref{sec::jagp_lr_space}).
We then compare this space to those of RHF (Section \ref{sec::compare_cis}) and CCSD (Section \ref{sec::compare_ccsd})
to help illuminate the differences between EOM-JAGP, CIS, and EOM-CCSD, after which we
conclude our theoretical analysis by discussing the origin and prevalence of ground-state bias in these methods
(Section \ref{sec::gs_bias}).
Results are presented for singlet and triplet excitations in
LiH (Section \ref{sec::results:lih}),
H$_2$O (Section \ref{sec::results:h2o}),
C$_2$ (Section \ref{sec::results:c2}),
and a collection of other small molecules (Section \ref{sec::results:other}).
We conclude (Section \ref{sec:conclusions}) with a summary of our findings and comments on the future development of EOM-VMC. 

\section{Theory}
\label{sec:theory}

\subsection{EOM Linear Response}
\label{sec::eom}

The general equation-of-motion (EOM) linear response formalism for an approximate ansatz may be derived by
assuming that the effects of a time-dependent perturbation to the Hamiltonian
may be accounted for by adding time-dependent perturbations to the wave function's variational parameters.
We therefore assume a Hamiltonian and wave function of the form
\begin{align}
\label{eqn:eom_ham}
H & = H_0 + e^{ -i \omega t } H_1 \\
\label{eqn:eom_wfn}
\left| \Psi \right> & = e^{ -i E_0 t / \hbar } \left| \Psi\left(\vec{x}(t)\right) \right> \\
\label{eqn:eom_vars}
\vec{x}(t) & = \vec{x}_0 + e^{ -i \omega t } \vec{\mu} \\
\label{eqn:eom_grnd_en}
E_0 & = \left< \Psi\left( \vec{x}_0 \right) \right| H_0 \left| \Psi \left( \vec{x}_0 \right) \right>
\end{align}
where $H_1$ and $\vec{\mu}$ are assumed to be small, $\vec{x}_0$ are the variational parameter values
for this ansatz's estimate of the unperturbed ground state, and $E_0$ is the corresponding estimate of the ground state energy.
Using the shorthand notation
\begin{align}
\label{eqn:shorthand_p0}
\left|\Psi^0\right> & \equiv \left|\Psi\left(\vec{x}_0\right)\right> \\
\label{eqn:shorthand_pi}
\left|\Psi^k\right> & \equiv \left[\partial \left|\Psi\left(\vec{x}\right)\right> / \partial x_k\right]_{\vec{x} \rightarrow \vec{x}_0}
\quad \quad (\mathrm{for}\hspace{2mm}k>0)
\end{align}
this wave function may be Taylor-expanded as
\begin{align}
\label{eqn:eom_talor}
\left| \Psi  \right>
& = e^{ -i E_0 t / \hbar } \left(
                                \left| \Psi^0 \right>
                              + e^{ -i \omega t } \sum_{k}^{}{ \mu_k \left| \Psi^k \right> }
                              + \mathcal{O}\left( |\vec{\mu}|^2 \right)
                           \right).
\end{align}
This expansion may be inserted into the Schr\"{o}dinger equation $i\hbar\partial|\Psi\rangle/\partial t=H|\Psi\rangle$ to give
\begin{align}
& H_1 \left|\Psi^0\right> + ( H_0 - E ) \sum_{k}^{}{ \mu_k \left|\Psi^k\right> } \notag \\
& \quad = e^{i \omega t} \left(E_0-H_0\right) \left|\Psi^0\right>
\label{eqn:eom_tdse}
\end{align}
where $E \equiv \hbar \omega + E_0$ and we have dropped terms quadratic in the perturbation or smaller.
Assuming the ground state variational principle is satisfied, and thus
\begin{align}
\label{eqn:gs_var}
\left< \Psi^k  \right| \left( H_0 - E_0 \right) \left| \Psi^0 \right> & = 0,
\end{align}
we may project Eq.\ (\ref{eqn:eom_tdse}) into the span of the ansatz's first derivatives
(i.e.\ left-project by $\langle\Psi^j|$)
to obtain
\begin{align}
\label{eqn:fds_proj}
  \sum_{k}^{}{ \left<\Psi^j\right| ( H_0 - E ) \left|\Psi^k\right> \mu_k }
= - \left<\Psi^j\right| H_1 \left|\Psi^0\right>.
\end{align}
Eq.\ (\ref{eqn:fds_proj}) is the EOM approximation for the response $\vec{\mu}$ to a small
perturbation $\exp(-i\omega t)H_1$ to the Hamiltonian.
Note that in particular, this approximation gives the resonances, i.e.\ the frequencies $\omega$
at which the response may be large even for a small perturbation, as the eigenvalues of
the unperturbed Hamiltonian $H_0$ in the subspace of Hilbert space spanned by the ansatz's first
derivatives with respect to its variational parameters.
In conclusion, obtaining the EOM estimates of the excitation energies $\hbar\omega=E-E_0$
therefore requires only that the Hamiltonian be diagonalized in this subspace by solving
\begin{align}
\label{eqn:eom_eig_eqn}
  \sum_{k}^{}{ \langle \Psi^j | H_0 | \Psi^k \rangle \mu_k }
= E \sum_{k}^{}{ \langle \Psi^j | \Psi^k \rangle \mu_k }.
\end{align}

\subsection{EOM-VMC}
\label{sec::theory_HS}

While deterministic methods exist to solve Eq.\ (\ref{eqn:eom_eig_eqn}) for RHF and CCSD (giving the CIS and EOM-CCSD methods, respectively), a stochastic
approach is more efficient in the case of wave functions with Jastrow factors like the JAGP.
Happily, the matrices involved are already available in many QMC software packages as they are the same matrices required for the ground state LM.
Here we briefly review how these matrices are estimated stochastically, and also point out some potential pitfalls when generalizing the methodology
for use in EOM.


To obtain our stochastic estimate to Eq.\ (\ref{eqn:eom_eig_eqn}), we first
insert resolutions of the identity, either in Fock space or real space,
\begin{align}
\label{eqn:roi_hilbert}
I & = \sum_{\vec{n}} \left|\vec{n}\right> \frac{|\left<\vec{n}|\Phi\right>|^2}{|\left<\vec{n}|\Phi\right>|^2} \left<\vec{n}\right| \\
\label{eqn:roi_real}
I & = \int d\vec{r} \left|\vec{r}\right> \frac{|\left<\vec{r}|\Phi\right>|^2}{|\left<\vec{r}|\Phi\right>|^2} \left<\vec{r}\right|
\end{align}
on both sides (note we will work in Fock space but the approach is equally well defined in real space) to obtain
\begin{align}
\label{eqn:eom_jagp_eigen_ri}
\begin{split}
&\sum_{\vec{n}} \sum_{j} \frac{ |\left< { \vec{n} }|{ { \Phi  } } \right>|^2 }{ \left< { { \Phi  } }|{ { \Phi  } } \right>  }
                         \frac{ |\left< { \vec{n} }|{ { \Psi^0} } \right>|^2 }{ |\left< { \vec{n} }|{ { \Phi  } } \right>|^2 }
                         \hspace{1.4mm} \mathcal{D}_{\vec{n},i}^{*} \hspace{1.4mm} \mathcal{G}_{\vec{n},j} \hspace{1.4mm} \mu_j
                         \\
& \quad = E
 \sum_{\vec{n}} \sum_{j} \frac{ |\left< { \vec{n} }|{ { \Phi  } } \right>|^2 }{ \left< { { \Phi  } }|{ { \Phi  } } \right>  }
                           \frac{ |\left< { \vec{n} }|{ { \Psi^0} } \right>|^2 }{ |\left< { \vec{n} }|{ { \Phi  } } \right>|^2 }
                           \hspace{1.4mm} \mathcal{D}_{\vec{n},i}^{*} \hspace{1.4mm} \mathcal{D}_{\vec{n},j} \hspace{1.4mm} \mu_j
\end{split}
\end{align}
where
\begin{align}
\label{eqn:lm_p_der_vecs}
\mathcal{D}_{\vec{n},j} & \equiv \frac { \left< { \vec{n} }|{ { \Psi  }^{ j } } \right>  }{ \left< { \vec{n} }|{ { \Psi^0  } } \right>  } \\
\label{eqn:lm_e_der_vecs}
\mathcal{G}_{\vec{n},j} & \equiv \frac { \left< { \vec{n} }|H_0|{ { \Psi  }^{ j } } \right>  }{ \left< { \vec{n} }|{ { \Psi^0  } } \right>  }
\end{align}
and we have introduced the importance sampling function $|\Phi\rangle$.
Eq.\ (\ref{eqn:eom_jagp_eigen_ri}) may be evaluated stochastically by an average on the Monte Carlo sample $\Omega$ drawn from $\Phi$'s
probability distribution by a Metropolis-Hastings walk, yielding
\begin{align}
\label{eqn:eom_jagp_eigen_mc}
\begin{split}
&\sum_{\vec{n}\in\Omega} \sum_{j} 
                         \frac{ |\left< { \vec{n} }|{ { \Psi^0} } \right>|^2 }{ |\left< { \vec{n} }|{ { \Phi  } } \right>|^2 }
                         \hspace{1.4mm} \mathcal{D}_{\vec{n},i}^{*} \hspace{1.4mm} \mathcal{G}_{\vec{n},j} \hspace{1.4mm} \mu_j
                         \\
& \quad = E
 \sum_{\vec{n}\in\Omega} \sum_{j} 
                           \frac{ |\left< { \vec{n} }|{ { \Psi^0} } \right>|^2 }{ |\left< { \vec{n} }|{ { \Phi  } } \right>|^2 }
                           \hspace{1.4mm} \mathcal{D}_{\vec{n},i}^{*} \hspace{1.4mm} \mathcal{D}_{\vec{n},j} \hspace{1.4mm} \mu_j
\end{split}
\end{align}
Thus, so long as a reasonable guiding function $\Phi$ is known and the ratios $\mathcal{D}_{\vec{n},j}$ and $\mathcal{G}_{\vec{n},j}$
can be evaluated efficiently, as is possible for the JAGP \cite{Neuscamman:2013:hilbert_jagp}, then EOM estimates of the excitation
energies may be evaluated for a cost similar to a ground state LM calculation.


However, while the ground state LM often makes the choice $|\Phi\rangle=|\Psi^0\rangle$, such a choice can be pathological in EOM-VMC
due to a ground state sampling bias.
To make this issue clear, consider the following simple model. 
Suppose we have a three-level system with $\left |1 \right> $,  $\left| 2 \right> $, and $\left| 3 \right> $ being its exact eigenstates
and we take as our ansatz the full configuration interaction (FCI) wave function, an arbitrary linear combination of all three states.
Next assume that we have already optimized the ground state perfectly, so $|\Psi^0\rangle=\left |1 \right>$.
Choosing $|\Phi\rangle=|\Psi^0\rangle$ would in this case prevent us from sampling the excited states at all, and 
so our stochastic estimate for the eigenvalue equation,
\begin{align}
\label{eqn:H_S}
&
\begin{pmatrix} 0 & 0 \\ 0 & 0 \end{pmatrix}
\begin{pmatrix} \mu_2 \\ \mu_3 \end{pmatrix}
= E
\begin{pmatrix} 0 & 0 \\ 0 & 0 \end{pmatrix}
\begin{pmatrix} \mu_2 \\ \mu_3 \end{pmatrix},
\end{align}
would be useless.
While in practice our ground state estimate $|\Psi^0\rangle$ is unlikely to be exact, using it as the guiding function will be statistically
inefficient, especially in cases when an excited state has a different symmetry than the ground state.


In future work it may be profitable to test general solutions to this problem.
In the present study, we have avoided sampling pathologies by adding random noise to the ground state JAGP's pairing matrix.
We find that noise distributed uniformly between -0.1 and 0.1 (when the largest pairing matrix element is 1) is effective.

\subsection{The JAGP Ansatz}
\label{sec::jagp}

Although the EOM-VMC formalism is applicable to any ansatz amenable to the ground state LM, we will limit our investigation in
this study to the Hilbert-space JAGP,
%
\begin{align}
\label{eqn:jagp}
&\left| { \Psi  }_{ \mathrm{JAGP} } \right> = \exp\left( \hat { J }  \right) \left| { \Psi  }_{ \mathrm{AGP} } \right> \\
\label{eqn:agp}
&\left| { \Psi  }_{ \mathrm{AGP} } \right> 
   ={ \left( \sum _{ r\overline { s }  } { { F }_{ r\overline { s }  }{ a }_{ r }^{ \dagger  }{ a }_{ \overline { s }  }^{ \dagger  } }  \right)  }^{ N/2 }\left| 0 \right> \\
\label{eqn:jastrow}
&\hat { J } 
   =\sum _{ p\le q }^{  }{ { J }_{ pq }^{ \alpha \alpha  }{ \hat { n }  }_{ p }{ \hat { n }  }_{ q } } +\sum _{ \overline { p } \le \overline { q }  }^{  }{ { J }_{ \overline { p } \overline { q }  }^{ \beta \beta  }{ \hat { n }  }_{ \overline { p }  }{ \hat { n }  }_{ \overline { q }  } } +\sum _{ p\overline { q }  }^{  }{ { J }_{ p\overline { q }  }^{ \alpha \beta  }{ \hat { n }  }_{ p }{ \hat { n }  }_{ \overline { q }  } }
\end{align}
where $N/2$ is the number of $\alpha$ (and $\beta$) electrons, unbarred and barred indices represent alpha and beta orbitals, respectively,
${\hat{n}}_{p}$ and ${a}_{p}^{\dagger}$ are the number and creation operator for the $p$th $\alpha$ orbital in the orthonormal orbital basis, and $\left|0\right>$ is the vacuum. 

While the JAGP has produced highly accurate results in a number of difficult molecules, especially upon optimization of its orbital basis
\cite{Neuscamman:2013:or_jagp,Neuscamman:2016:lm_cjagp},
such results have all been obtained through non-linear parameter optimizations targeted at individual eigenstates.
In the present study, we seek to determine its efficacy when such individual state optimizations are eschewed in favor of an EOM approach.
To better understand what capabilities and limitations to expect in this new use of JAGP, we will discuss the nature of its linear response space
and make formal comparisons to other EOM methods, namely CIS and EOM-CCSD.


\subsection{JAGP's Linear Response Space}
\label{sec::jagp_lr_space}

The accuracy of any LR based methods depend on both the number and nature of the ansatz's first derivatives.
In considering the nature of the JAGP's wave function derivatives, we separate them into those for the pairing matrix (F) and Jastrow factor (J)
variables. 

The AGP by itself is able to create closed-shell and open-shell configurations.
Consider the simple H$_2$ molecule in a minimal basis, noting that rotations of the one-particle basis
will not change the span of the AGP's first derivatives and that we can analyze its properties under any rotation that is convenient.
If we work in molecular orbital basis, for example, and label the bonding orbital as 1 and anti-bonding
orbital as 2, then the RHF solution, encoded in an AGP pairing matrix, is:
\begin{align}
\label{eqn:h2_agp1}
F_{\mathrm{closed-shell}} = \begin{pmatrix} 1 & 0 \\ 0 & 0 \end{pmatrix}
\end{align}
in which the nonzero matrix element creates a pair of electrons in the bonding orbital. 
Similarly, the HOMO$\rightarrow$ LUMO singlet open-shell configuration is encoded as:
\begin{align}
\label{eqn:h2_agp2}
F_{\mathrm{open-shell}} = \begin{pmatrix} 0 & 1/\sqrt { 2 }  \\ 1/\sqrt { 2 }  & 0 \end{pmatrix}
\end{align}
Clearly, the open-shell pairing matrix can be written as a sum of derivatives of the closed-shell matrix with respect to its elements,
and so we would expect EOM-AGP, like CIS, to be capable of describing this type of simple excitation.
%

The derivative of the JAGP with respect to a pairing matrix element is:
\begin{align}
\label{eqn:pm_der}
\frac { \partial \left| { \Psi  }_{ JAGP } \right>  }{ \partial { F }_{ p\overline { q }  } } 
   =\left( N/2 \right) exp\left( \hat { J }  \right) { a }_{ p }^{ \dagger  }{ a }_{ \overline { q }  }^{ \dagger  }{ \left( \sum _{ r\overline { s }  }^{  }{ { F }_{ r\overline { s }  }{ a }_{ r }^{ \dagger  } } { a }_{ \overline { s }  }^{ \dagger  } \right)  }^{ \left( N/2-1 \right)  }
\end{align}
which, although more complicated than the simple $H_2$ example, will have a similar physical meaning when the ground state is dominated
by the RHF determinant.
In this discussion we will limit our analysis to this single-reference case, although it would be interesting in future to investigate how
the derivatives change in more multi-configurational settings.

If $p$ and $\overline{q}$ are both occupied in the ground state,
the derivative above will essentially give the ground state wave function back and little information will be gained about excited states.
If $p$ is occupied but $\overline{q}$ is empty in the ground state, 
this derivative will create a $p\rightarrow\overline{q}$ single excitation.
If both $p$ and $\overline{q}$ are empty in the ground state, this derivative will create a double excitation.

Derivatives with respect to Jastrow factor variables,
\begin{align}
\label{eqn:jas_der}
\frac { \partial \left| { \Psi  }_{ JAGP } \right>  }{ \partial { J }_{ p\overline { q }  }^{ \alpha \beta  } } 
   ={ \hat { n }  }_{ p }{ \hat { n }  }_{ \overline { q }  }\left| { \Psi  }_{ JAGP } \right>
\end{align}
although easy to evaluate, are not so easily analyzed as those for the pairing matrix, in part because their
character is strongly dependent on the one-particle basis chosen for the number operators.
In this work, as in other studies of the Hilbert-space JAGP, this basis is chosen to be local.
Thus Jastrow derivatives produce projections of the ground state wave function in which two particular local
orbitals are guaranteed to be occupied.
While many such projections are no doubt components of excited states, predicting their significance in EOM-JAGP
is not so straightforward.
At best, the coupled cluster interpretation \cite{Neuscamman:2013:or_jagp} of the Jastrow factor would suggest 
that such derivatives provide a limited subset of the excitations present in EOM-CCSD.
However, this subset will have been optimized for the purposes of lowering the ground state energy,
and as it is only a small subset, it would be a surprise if it could reproduce the highly flexible linear response
space provided by the coupled cluster doubles operator.

%
%

\subsection{Comparison with CIS}
\label{sec::compare_cis}

CIS, equivalent to EOM-RHF, has a first derivative subspace of size $N_{occ}N_{vir}$, consisting exclusively of single excitations out of the
RHF determinant.
Thus both EOM-JAGP and CIS have a first derivative space of size $\mathcal{O}\left(N^2\right)$, with EOM-JAGP's being larger by a constant prefactor.
Given that it has a larger EOM subspace, contains RHF as a special case, and has some potential for treating double excitations, one might
expect EOM-JAGP to be strictly superior to CIS in terms of accuracy in excitation energies.
While this appears to be true in our results for relative excitation energies, it is not always true for absolute excitation energies due to
JAGP's much stronger ground state bias (see Section \ref{sec::gs_bias}).

\subsection{Comparison with EOM-CCSD}
\label{sec::compare_ccsd}

Like EOM-CCSD, EOM-JAGP has the potential to treat both single and double excitations.
Further, given the presence of double excitations in its EOM subspace, one might expect JAGP
to benefit from the tendency, common in EOM-CCSD, of these excitations to act to relax the wave function's
orbitals in the presence of a single excitation.
%
However, the double excitations in EOM-JAGP are much less flexible than in EOM-CCSD, a reality made clear by
a close look at the the CCSD wave function:
\begin{align}
\label{eqn:ccsd_ansatz}
\begin{split}
&\left| \Psi_{\mathrm{CCSD}} \right> = \exp\left( { \hat { T }  }_{ 1 }+{ \hat { T }  }_{ 2 } \right) \left| RHF \right> \\
&{ \hat { T }  }_{ 1 }=\sum _{ ia }^{  }{ { t }_{ i }^{ a }{ a }_{ i }{ a }_{ a }^{ \dagger  } } \\
&{ \hat { T }  }_{ 2 }=\sum _{ i>j,a>b }^{  }{ { t }_{ ij }^{ ab }{ a }_{ i }{ a }_{ a }^{ \dagger  }{ a }_{ j }{ a }_{ b }^{ \dagger  } } \\
\end{split}
\end{align}
From this expression, we can see that derivatives with respect to the cluster amplitudes will produce an EOM subspace containing $N_{occ}N_{vir}$
single excitations and $\mathcal{O}\left(N_{occ}^2N_{vir}^2\right)$ double excitations.
Contrast this with EOM-JAGP, where we find only $\mathcal{O}\left(N_{vir}^2\right)$ double excitations, suggesting a great disparity in flexibility with respect
to doubles.
Beyond sheer number, the occupied-orbital indexation of the CCSD doubles gives EOM-CCSD direct control of which occupied orbitals a double excitation is promoted
from, whereas the double excitations in EOM-JAGP are in effect indexed only by their virtual orbitals, creating what we think of as the ``uncontrolled hole''
problem in which EOM-JAGP has difficulty ensuring that a double excitation be promoted from physically reasonable occupied orbitals.
%
%
In practice, we will see that these disparities prevent EOM-JAGP from achieving the high accuracy typical of EOM-CCSD, presumably because they rob it of the ability
to carefully relax orbital shapes in the presence of an excitation, although interestingly their performance is more similar (although not particularly good) 
for double excitations.

\subsection{Ground State Bias}
\label{sec::gs_bias}

All EOM-based methods should be expected to suffer from a bias in favor of the ground state, and therefore too-high excitation energies,
as the initial variational parameters $\vec{x}_0$ have been optimized for this state in a nonlinear fashion that takes into
account interactions between the effects of different parameters.
Linear response methods, by their very nature, cannot achieve this degree of tailoring for the excited states.
For example, in CIS, the EOM subspace contains the freedom to shape the excitation's orbital, but cannot achieve the second-order
effect of relaxing the shapes of other orbitals in the presence of the excitation.
This lack of orbital relaxation is perhaps in practice the most common and important source of ground state bias, showing up also
in EOM-CCSD in the case of doubly excited states (for which EOM-CCSD has no triples to use to couple in relaxation in the way
it can via its doubles for singly excited states).

In EOM-JAGP we find that ground state bias can be particularly severe, because while it's EOM subspace is closer in its flexibility
to that of CIS than to that of EOM-CCSD, the correlation included by and thus the accuracy of ground state JAGP is closer to that of CCSD.
From the EOM perspective, JAGP is in a sense too clever for its own good:  by capturing a large amount of ground state correlation
energy using an ansatz with a small number of parameters and thus a relatively inflexible EOM subspace, it is virtually guaranteed
to have a sizable bias in favor of the ground state.
As we will see, this bias is severe enough that it tends to overestimate excitation energies even more so than CIS, despite having
a somewhat more flexible EOM space.
Fortunately, as the ground state bias affects all excitation energies roughly equally, relative energies between different excitations
should be little affected and should be expected to show improvement over methods (like CIS) that have less flexible EOM spaces.

While it is easy to confirm the presence of ground state bias by comparing absolute excitation energies to those of a benchmark
method, analyzing the accuracy of relative excitation energies is less straightforward.
The approach we take rests on the idea that two methods, both with exact relative energies between
excited states but with different ground state biases, can be made the same by applying a constant shift, and thus we seek
a measure of relative accuracy that will automatically account for any such constant shift.
To this effect we will use a root-mean-square relative deviation (RMSRD) metric
\begin{align}
\label{eqn:rmsrdn}
&\mathrm{RMSRD}_N \equiv
\left( \frac{1}{N} \sum_i^N \left[ \Delta_{i,\mathrm{Method}}  - \Delta_{i,\mathrm{Benchmark}} \right]^2 \right)^{ 1/2 }
\end{align}
where $\Delta_i=\hbar(\omega_i - \bar{\omega})$ is the deviation of a method's $i$th excitation energy ($\hbar\omega_i$) from
the mean ($\hbar\bar{\omega}$) of that method's first $N$ excitation energies.
RMSRD$_N$ thus measures how closely a method's excitation energies' deviations from their own mean match the corresponding
deviations in a benchmark method, and so the RMSRD$_N$ for a method with exact relative energies between excited states but
a large ground state bias would be zero, while that for a method with no ground state bias overall but large errors in relative
excitation energies would be large.
Thus, by analyzing both absolute excitation energies and excitation energies' deviations from their own mean, we will attempt
to distinguish the effect of ground state bias from other sources of error.

\section{Results}
\label{sec:results}

\subsection{Computational Details}
\label{sec::comp_details}

Before we present our results, let us briefly overview the computational details.
EOM-CCSD, Davidson-corrected MRCI (MRCI+Q) and FCI results were computed with MOLPRO
\cite{MOLPRO_brief},
CIS results with QChem \cite{Qchem:2013}, and JAGP results with our own prototype Hilbert space quantum Monte Carlo code
with one- and two-electron integrals imported from Psi3 \cite{Psi3}.
In JAGP, we worked exclusively in the symmetrically orthogonalized $S^{-1/2}$ orbital basis and froze the C, N and O 1s orbitals
at the RHF level.
Unless noted otherwise, all sample lengths were $7.2\times { 10 }^{ 6 }$.
All statistical uncertainties were converged to less than 0.01eV in all cases. 

\subsection{Case Study 1: LiH in cc-pVDZ}
\label{sec::results:lih}

\begin{figure}
 \includegraphics[width=9.0cm, angle=0]{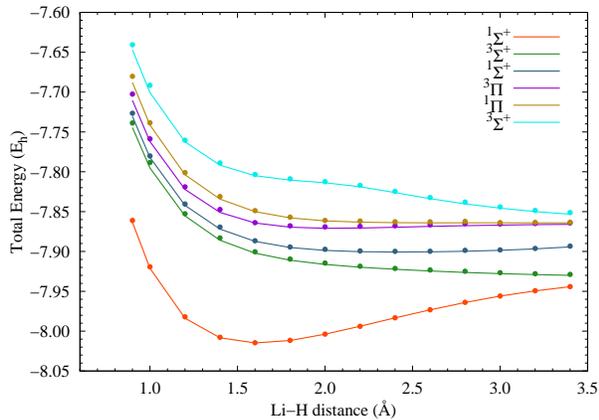}
 \caption{Potential energy curves for the lowest excited states of lithium hydride in the cc-pVDZ basis set.
          Solid lines are FCI while dots are EOM-JAGP calculation.
          EOM-JAGP correctly finds the lowest three triplet and two singlet states.}
 \label{fig:LiH}
\end{figure}

We begin our results with a simple example, the LiH molecule in a cc-pVDZ basis \cite{cc-pvdz}.
This system has only two valence electrons, and thus we expect the two-electron nature of the AGP to result
in high accuracy for EOM-JAGP.
Due to the two-electron nature of AGP and the fact that all low lying excited states are singly excited states,
we expect nearly exact results.
Figure \ref{fig:LiH} shows our results for the lowest five excited states at 14 bond lengths between 0.9 $\mathring { A } $ and 3.4 $\mathring { A } $. 
As expected, EOM-JAGP is quite accurate, with an overall average error from FCI of just 2 milliHartree. 




\subsection{Case Study 2: H$_2$O in 6-31G}
\label{sec::results:h2o}

\begin{table}[t]
\centering
\caption{Vertical excitation energies and RMSRD$_5$ values (both in eV) for
         the water molecule at equilibrium in a 6-31G basis set.
        }
\label{tab:water}
\begin{tabular}{  c  r@{.}l  r@{.}l  r@{.}l  r@{.}l  }
\hline\hline
\multicolumn{1}{ c }{ \hspace{0mm}   \hspace{0mm} } &
\multicolumn{8}{ c }{ \hspace{0mm}  Excitation Energy (eV)  \hspace{0mm} } \\
\hline
 \hspace{0mm} State \hspace{0mm} &
 \multicolumn{2}{ c }{ \hspace{0mm}  EOM-JAGP       \hspace{0mm} } &
 \multicolumn{2}{ c }{ \hspace{0mm}  CIS       \hspace{0mm} } &
 \multicolumn{2}{ c }{ \hspace{0mm}  EOM-CCSD       \hspace{0mm} } &
 \multicolumn{2}{ c }{ \hspace{0mm}  FCI       \hspace{0mm} } \\
\hline
 \hspace{0mm}  $^{3}B_2$   \hspace{0mm}  & \hspace{0mm} 6&95 \hspace{0mm} & \hspace{0mm} 6&64 \hspace{0mm} & \hspace{0mm} 6&03 \hspace{0mm} & \hspace{0mm} 6&10 \hspace{0mm} \\
 \hspace{0mm}  $^{1}B_1$   \hspace{0mm}  & \hspace{0mm} 7&89 \hspace{0mm} & \hspace{0mm} 7&71 \hspace{0mm} & \hspace{0mm} 6&79 \hspace{0mm} & \hspace{0mm} 6&85 \hspace{0mm} \\
 \hspace{0mm}  $^{3}A_1$   \hspace{0mm}  & \hspace{0mm} 8&39 \hspace{0mm} & \hspace{0mm} 7&82 \hspace{0mm} & \hspace{0mm} 8&02 \hspace{0mm} & \hspace{0mm} 8&07 \hspace{0mm} \\
 \hspace{0mm}  $^{3}A_2$   \hspace{0mm}  & \hspace{0mm} 9&04 \hspace{0mm} & \hspace{0mm} 8&51 \hspace{0mm} & \hspace{0mm} 8&17 \hspace{0mm} & \hspace{0mm} 8&17 \hspace{0mm} \\
 \hspace{0mm}  $^{1}A_1$   \hspace{0mm}  & \hspace{0mm} 9&73 \hspace{0mm} & \hspace{0mm} 9&31 \hspace{0mm} & \hspace{0mm} 8&74 \hspace{0mm} & \hspace{0mm} 8&75 \hspace{0mm} \\
\hline\hline
 \hspace{0mm}  RMSRD$_5$   \hspace{0mm}  & \hspace{0mm} 0&26 \hspace{0mm} & \hspace{0mm} 0&37 \hspace{0mm} & \hspace{0mm} 0&02 \hspace{0mm} & \hspace{0mm} 0&00 \hspace{0mm} \\
\hline\hline
\end{tabular}
\end{table}

\begin{figure}
 \includegraphics[width=8.0cm, angle=270]{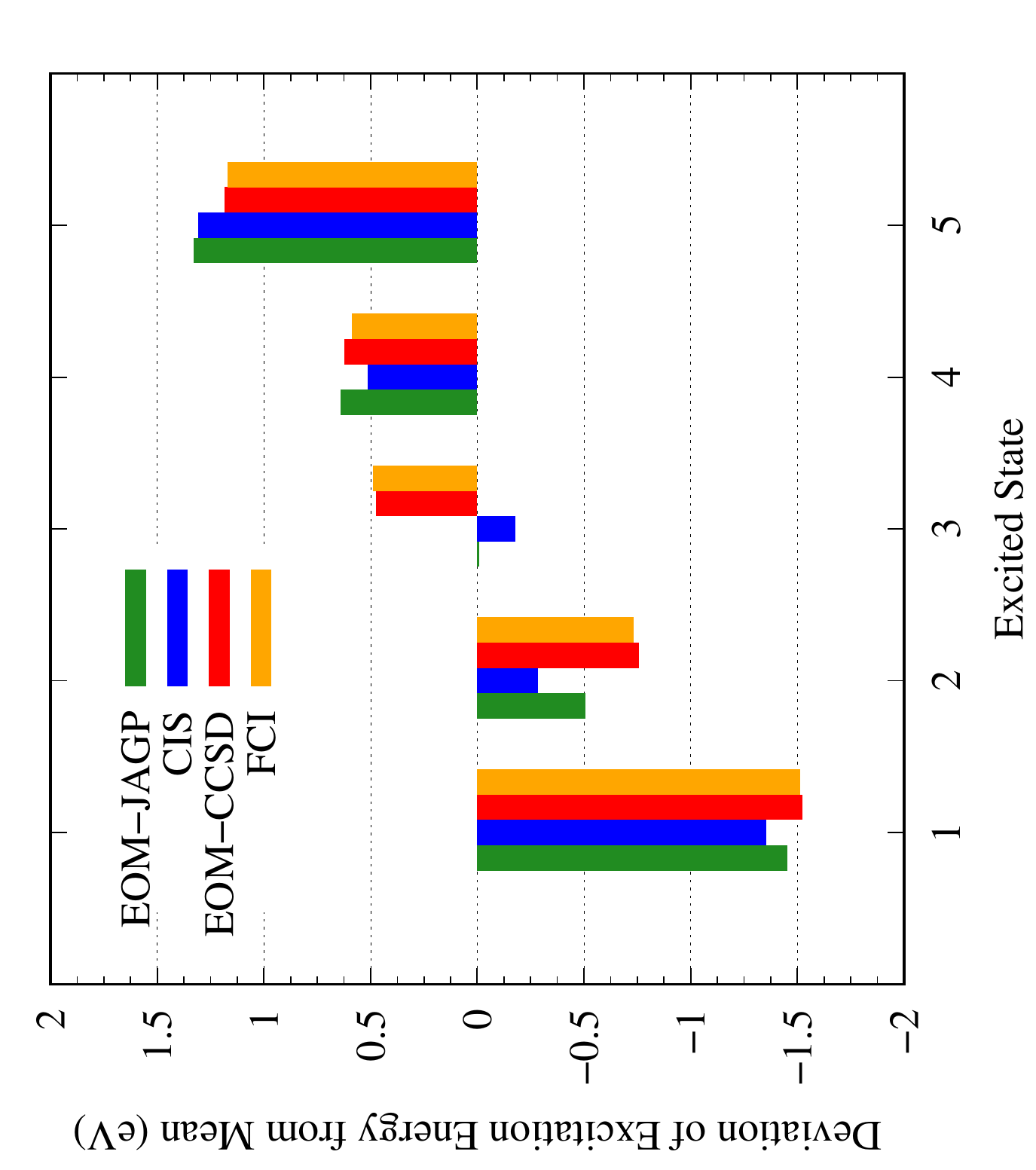}
 \caption{For each method, the deviation of each excitation energy from that method's mean excitation energy for H$_2$O in a 6-31G basis.}
 \label{fig:H2O}
\end{figure}

The water molecule provides an excellent illustration of how the different EOM methods we discuss are expected to perform for single excitations
in a single reference system.
As is well known, EOM-CCSD proves highly accurate in this setting.
EOM-JAGP and CIS are less accurate, but in different ways.
Table \ref{tab:water} shows that EOM-JAGP tends to overestimate water's excitation energies, even more so than does CIS, in keeping with
the expectation of a more severe ground state bias.
Also as expected, the RMSRD$_5$ values and Figure \ref{fig:H2O} show that EOM-JAGP produces more accurate relative energies between excitations than CIS,
but less accurate relative energies than EOM-CCSD.

\subsection{Case Study 3: C$_2$ in 6-31G}
\label{sec::results:c2}

\begin{table}[t]
\centering
\caption{Vertical excitation energies and RMSRD$_5$ values (both in eV) for C$_2$ at equilibrium in a 6-31G basis set.}
\label{tab:c2}
\begin{tabular}{  c  r@{.}l  r@{.}l  r@{.}l  r@{.}l  }
\hline\hline
\multicolumn{1}{ c }{ \hspace{0mm}   \hspace{0mm} } &
\multicolumn{8}{ c }{ \hspace{0mm}  Excitation Energy       \hspace{0mm} } \\
\hline
 \hspace{0mm} State \hspace{0mm} &
 \multicolumn{2}{ c }{ \hspace{0mm}  EOM-JAGP       \hspace{0mm} } &
 \multicolumn{2}{ c }{ \hspace{0mm}  CIS       \hspace{0mm} } &
 \multicolumn{2}{ c }{ \hspace{0mm}  EOM-CCSD       \hspace{0mm} } &
 \multicolumn{2}{ c }{ \hspace{0mm}  FCI       \hspace{0mm} } \\
\hline
 \hspace{0mm}  $^{ 3 }{ \Sigma  }_{ g }^{ - }$   \hspace{0mm}  & \hspace{0mm} 0&61 \hspace{0mm} & \hspace{0mm} -1&46 \hspace{0mm} & \hspace{0mm} 0&59 \hspace{0mm} & \hspace{0mm} 0&60 \hspace{0mm} \\
 \hspace{0mm}  $^{ 3 }{ \Pi  }_{ u }$   \hspace{0mm}  & \hspace{0mm} 1&12 \hspace{0mm} & \hspace{0mm} -2&14 \hspace{0mm} & \hspace{0mm} 0&88 \hspace{0mm} & \hspace{0mm} 0&89 \hspace{0mm} \\
 \hspace{0mm}  $^{ 3 }{ \Pi  }_{ u }$   \hspace{0mm}  & \hspace{0mm} 1&12 \hspace{0mm} & \hspace{0mm} -2&14 \hspace{0mm} & \hspace{0mm} 0&88 \hspace{0mm} & \hspace{0mm} 0&89 \hspace{0mm} \\
 \hspace{0mm}  $^{ 1 }{ \Pi  }_{ u }$   \hspace{0mm}  & \hspace{0mm} 2&41 \hspace{0mm} & \hspace{0mm} -0&98 \hspace{0mm} & \hspace{0mm} 2&36 \hspace{0mm} & \hspace{0mm} 2&17 \hspace{0mm} \\
 \hspace{0mm}  $^{ 1 }{ \Pi  }_{ u }$   \hspace{0mm}  & \hspace{0mm} 2&41 \hspace{0mm} & \hspace{0mm} -0&98 \hspace{0mm} & \hspace{0mm} 2&36 \hspace{0mm} & \hspace{0mm} 2&17 \hspace{0mm} \\
 \hspace{0mm}  $^{ 1 }{ \Sigma  }_{ g }^{ + }$   \hspace{0mm}  & \hspace{0mm} 4&30 \hspace{0mm} & 
 \multicolumn{2}{ c }{ \hspace{0mm}  N/A    \hspace{0mm} } &
 \hspace{0mm} 4&39 \hspace{0mm} & 
 \hspace{0mm} 3&28 \hspace{0mm} \\
\hline\hline
 \hspace{0mm}  RMSRD$_5$   \hspace{0mm}  & \hspace{0mm} 0&09 \hspace{0mm} & \hspace{0mm} 0&42 \hspace{0mm} & \hspace{0mm} 0&09 \hspace{0mm} & \hspace{0mm} 0&00 \hspace{0mm} \\
\hline\hline
\end{tabular}
\end{table}

\begin{figure}
 \includegraphics[width=8.0cm, angle=270]{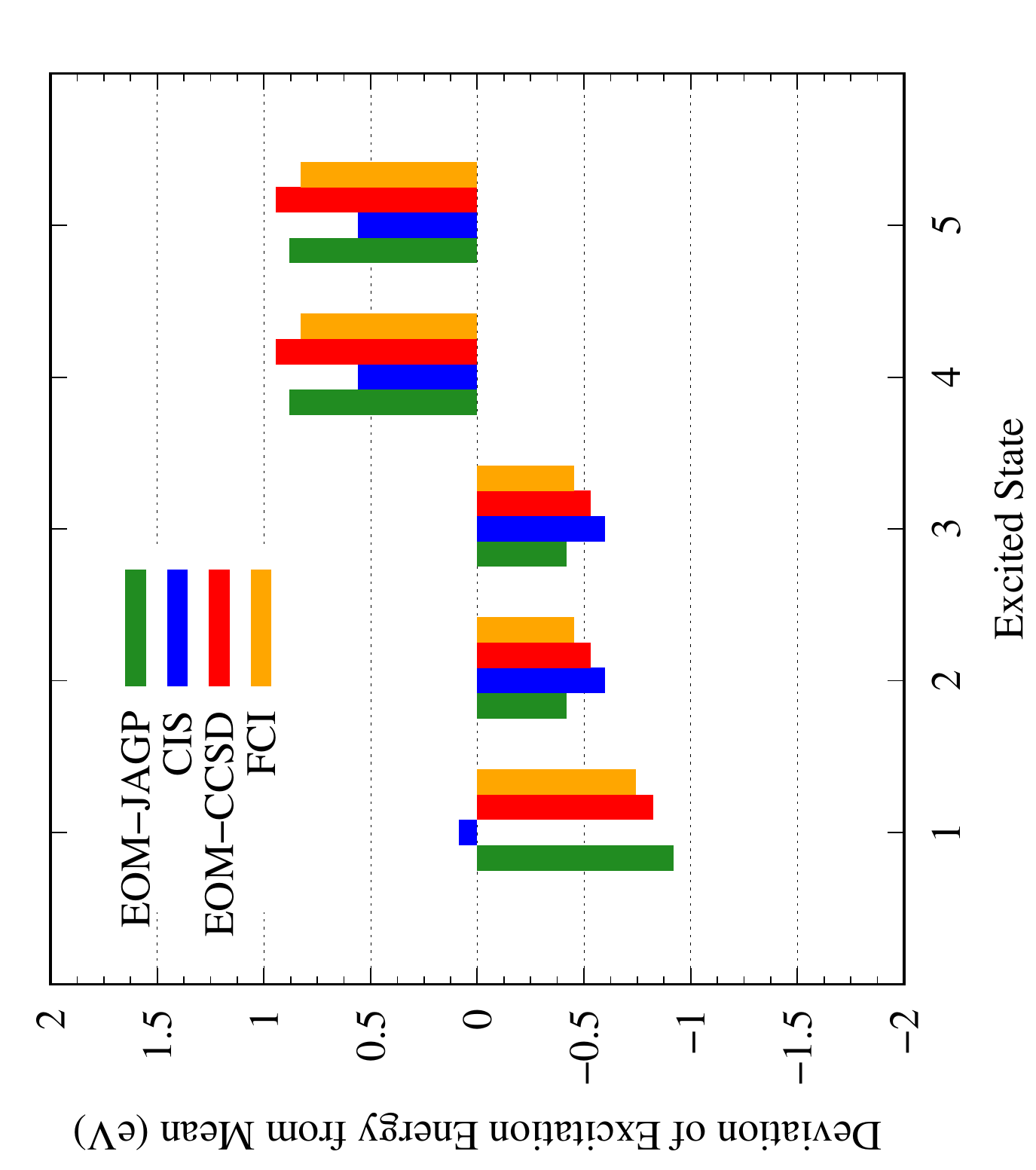}
 \caption{For each method, the deviation of each excitation energy from that method's mean excitation energy for C$_2$ in a 6-31G basis.}
 \label{fig:C2}
\end{figure}

Our last case study is C$_2$ molecule in 6-31G \cite{6-31g} basis set, which presents a major challenge for EOM methods due to
the presence of strong correlation in the ground state and low-lying doubly excited states.
Indeed, the very poor quality of RHF for the ground state of C$_2$ leads CIS to predict the first five excited states to lie \textit{below}
the ground state in energy, and its limitation to single excitations causes CIS to miss the doubly-excited sixth excited state entirely.
The presence of strong correlation is not nearly so problematic for EOM-JAGP, which performs if anything better than expected, especially in
absolute energies for single excitations (the first five excitations) in which it displays very little ground state bias.
The most likely explanation for this lack of bias lies in the ground state JAGP's inability to capture as high a fraction of the correlation energy
as in H$_2$O (although it is still vastly superior to RHF), and so some cancellation of error appears to be at work.
Overall, the EOM-JAGP results are similar to those of EOM-CCSD, being accurate for single excitations but having a much too high energy
for the doubly excited state due to a lack of triples excitations in their EOM subspaces.


\subsection{Other Benchmarking Calculations}
\label{sec::results:other}

\begin{figure}
 \includegraphics[width=9.0cm, angle=0]{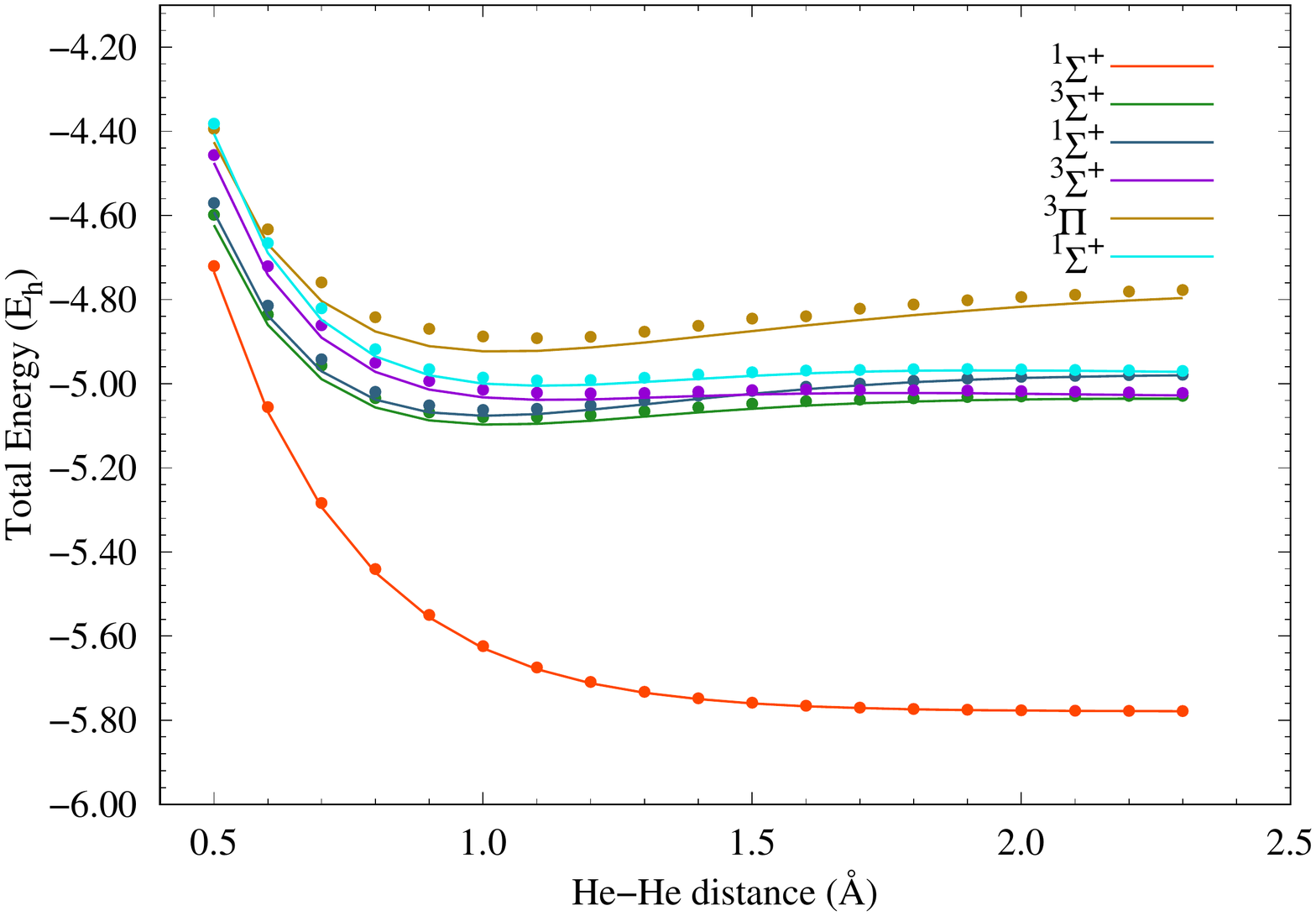}
 \caption{Potential energy curves for the lowest excited states of the helium dimer in the aug-cc-pVDZ \cite{aug-cc-pvdz} basis set.
          Solid lines correspond to FCI, dots to EOM-JAGP.
          The EOM-JAGP correctly finds the lowest three triplet and two singlet states.}
 \label{fig:He2}
\end{figure}

\begin{table}[t]
\centering
\caption{Vertical excitation energies and RMSRD$_N$ values (both in eV) for molecules at equilibrium in a 6-31G basis.}
\label{tab:diatomic}
\begin{tabular}{  c  c  r@{.}l  r@{.}l  r@{.}l  r@{.}l  }
\hline\hline
\multicolumn{2}{ c }{ \hspace{0mm}  \hspace{0mm} } &
\multicolumn{8}{ c }{ \hspace{0mm}  Excitation Energy    \hspace{0mm} } \\
\hline
 \hspace{0mm} Molecule \hspace{0mm} &
 \hspace{0mm} State \hspace{0mm} &
 \multicolumn{2}{ c }{ \hspace{0mm}  EOM-JAGP    \hspace{0mm} } &
 \multicolumn{2}{ c }{ \hspace{0mm}  CIS    \hspace{0mm} } &
 \multicolumn{2}{ c }{ \hspace{0mm}  EOM-CCSD    \hspace{0mm} } &
 \multicolumn{2}{ c }{ \hspace{0mm}  FCI    \hspace{0mm} } \\
\hline
 \hspace{0mm} Li$_2$  \hspace{0mm} &  $^{ 3 }{ \Sigma  }_{ u }^{ + }$   \hspace{0mm}  & \hspace{0mm} 1&20 \hspace{0mm} & \hspace{0mm} 0&63 \hspace{0mm} & \hspace{0mm} 1&18 \hspace{0mm} & \hspace{0mm} 1&17 \hspace{0mm} \\
 \hspace{0mm}         \hspace{0mm} &  $^{ 3 }{ \Pi  }_{ g }$   \hspace{0mm}  & \hspace{0mm} 1&45 \hspace{0mm} & \hspace{0mm} 0&86 \hspace{0mm} & \hspace{0mm} 1&42 \hspace{0mm} & \hspace{0mm} 1&42 \hspace{0mm}  \\
 \hspace{0mm}         \hspace{0mm} &  $^{ 1 }{ \Pi  }_{ g }$   \hspace{0mm}  & \hspace{0mm} 1&90 \hspace{0mm} & \hspace{0mm} 2&32 \hspace{0mm} & \hspace{0mm}  1&86 \hspace{0mm} & \hspace{0mm} 1&86 \hspace{0mm}  \\
 \hspace{0mm}         \hspace{0mm} &  $^{ 3 }{ \Sigma  }_{ u }^{ + }$   \hspace{0mm}  & \hspace{0mm} 2&27 \hspace{0mm} & \hspace{0mm} 1&67 \hspace{0mm} & \hspace{0mm} 2&23 \hspace{0mm} & \hspace{0mm} 2&23 \hspace{0mm}  \\
 \hspace{0mm}         \hspace{0mm} &  $^{ 1 }{ \Sigma  }_{ u }^{ + }$   \hspace{0mm}  & \hspace{0mm} 2&90 \hspace{0mm} & \hspace{0mm} 3&21 \hspace{0mm} & \hspace{0mm}  2&87 \hspace{0mm} & \hspace{0mm} 2&87 \hspace{0mm}  \\
 \hspace{0mm}         \hspace{0mm} &  RMSRD$_5$   \hspace{0mm} & \hspace{0mm} 0&00 \hspace{0mm} & \hspace{0mm} 0&47 \hspace{0mm} & \hspace{0mm} 0&00 \hspace{0mm} & \hspace{0mm} 0&00 \hspace{0mm} \\
\hline
 \hspace{0mm} N$_2$   \hspace{0mm} &  $^{ 3 }{ \Sigma  }_{ u }^{ - }$   \hspace{0mm}  & \hspace{0mm} 7&89 \hspace{0mm} & \hspace{0mm} 6&10 \hspace{0mm} & \hspace{0mm} 7&97 \hspace{0mm} & \hspace{0mm} 7&87 \hspace{0mm}  \\
 \hspace{0mm}         \hspace{0mm} &  $^{ 3 }{ \Pi  }_{ g }$   \hspace{0mm}  & \hspace{0mm} 8&81 \hspace{0mm} & \hspace{0mm} 7&31 \hspace{0mm} & \hspace{0mm} 8&01 \hspace{0mm} & \hspace{0mm} 7&94 \hspace{0mm}  \\
 \hspace{0mm}         \hspace{0mm} &  $^{ 3 }{ \Sigma  }_{ g }^{ + }$   \hspace{0mm}  & \hspace{0mm} 9&45 \hspace{0mm} & \hspace{0mm} 7&70 \hspace{0mm} & \hspace{0mm} 9&58 \hspace{0mm} & \hspace{0mm} 9&36 \hspace{0mm}  \\
 \hspace{0mm}         \hspace{0mm} &  RMSRD$_3$   \hspace{0mm} & \hspace{0mm} 0&39 \hspace{0mm} & \hspace{0mm} 0&52 \hspace{0mm} & \hspace{0mm} 0&06 \hspace{0mm} & \hspace{0mm} 0&00 \hspace{0mm} \\ 
\hline
 \hspace{0mm} CO  \hspace{0mm} &  $^{ 3 }{ \Pi  }$   \hspace{0mm}  & \hspace{0mm} 6&57 \hspace{0mm} & \hspace{0mm} 5&66 \hspace{0mm} & \hspace{0mm} 6&16 \hspace{0mm} & \hspace{0mm} 6&01 \hspace{0mm}  \\
 \hspace{0mm}         \hspace{0mm} &  $^{ 3 }{ \Sigma  }^{ - }$   \hspace{0mm}  & \hspace{0mm} 8&75 \hspace{0mm} & \hspace{0mm} 7&63 \hspace{0mm} & \hspace{0mm} 8&53 \hspace{0mm} & \hspace{0mm} 8&61 \hspace{0mm}  \\
 \hspace{0mm}         \hspace{0mm} &  $^{ 1 }{ \Pi  }$   \hspace{0mm}  & \hspace{0mm} 9&32 \hspace{0mm} & \hspace{0mm} 8&74 \hspace{0mm} & \hspace{0mm} 8&63 \hspace{0mm} & \hspace{0mm} 8&67 \hspace{0mm}  \\
 \hspace{0mm}         \hspace{0mm} &  RMSRD$_3$   \hspace{0mm} & \hspace{0mm} 0&22 \hspace{0mm} & \hspace{0mm} 0&43 \hspace{0mm} & \hspace{0mm} 0&10 \hspace{0mm} & \hspace{0mm} 0&00 \hspace{0mm} \\
\hline
 \hspace{0mm} CH$_2$  \hspace{0mm} &  $^{ 3 }{ B }_{ 1 }$   \hspace{0mm}  & \hspace{0mm} 0&26 \hspace{0mm} & \hspace{0mm} -0&74 \hspace{0mm} & \hspace{0mm} -0&38 \hspace{0mm} & \hspace{0mm} -0&38 \hspace{0mm}  \\
 \hspace{0mm}         \hspace{0mm} &  $^{ 1 }{ A }_{ 1 }$   \hspace{0mm}  & \hspace{0mm} 5&84 \hspace{0mm} &  
 \multicolumn{2}{ c }{ \hspace{0mm}  N/A    \hspace{0mm} } &
 \hspace{0mm} 5&84 \hspace{0mm} & \hspace{0mm} 4&81 \hspace{0mm} \\
 \hspace{0mm}         \hspace{0mm} &  RMSRD$_2$   \hspace{0mm} & \hspace{0mm} 0&20 \hspace{0mm} & 
 \multicolumn{2}{ c }{ \hspace{0mm}  N/A    \hspace{0mm} } &
 \hspace{0mm} 0&52 \hspace{0mm} & \hspace{0mm} 0&00 \hspace{0mm} \\
\hline\hline
\end{tabular}
\end{table}

\begin{table}[t]
\centering
\caption{Vertical excitation energies and RMSRD$_N$ values (both in eV) for molecules at equilibrium in a 6-31G basis.}
\label{tab:multiatomic}
\begin{tabular}{  c  c  r@{.}l  r@{.}l  r@{.}l  r@{.}l  }
\hline\hline
\multicolumn{2}{ c }{ \hspace{0mm}   \hspace{0mm} } &
\multicolumn{8}{ c }{ \hspace{0mm}  Excitation Energy       \hspace{0mm} } \\
\hline
 \hspace{0mm} Molecule \hspace{0mm} &
 \hspace{0mm} State \hspace{0mm} &
 \multicolumn{2}{ c }{ \hspace{0mm}  EOM-JAGP       \hspace{0mm} } &
 \multicolumn{2}{ c }{ \hspace{0mm}  CIS       \hspace{0mm} } &
 \multicolumn{2}{ c }{ \hspace{0mm}  EOM-CCSD       \hspace{0mm} } & 
 \multicolumn{2}{ c }{ \hspace{0mm}  MRCI+Q    \hspace{0mm} } \\
\hline
 \hspace{0mm} HCN  \hspace{0mm} &  $^{ 3 }{ \Sigma  }^{ - }$   \hspace{0mm}  & \hspace{0mm} 6&57 \hspace{0mm} & \hspace{0mm} 5&09 \hspace{0mm} & \hspace{0mm} 6&57 \hspace{0mm} & \hspace{0mm} 6&55 \hspace{0mm}  \\
 \hspace{0mm}         \hspace{0mm} &  $^{ 3 }{ \Sigma  }^{ - }$   \hspace{0mm}  & \hspace{0mm} 8&20 \hspace{0mm} & \hspace{0mm} 6&26 \hspace{0mm} & \hspace{0mm} 8&18 \hspace{0mm} & \hspace{0mm} 8&02 \hspace{0mm}  \\
 \hspace{0mm}         \hspace{0mm} &  $^{ 3 }{ \Sigma  }^{ - }$   \hspace{0mm}  & \hspace{0mm} 8&83 \hspace{0mm} & \hspace{0mm} 7&22 \hspace{0mm} & \hspace{0mm} 8&92 \hspace{0mm} & \hspace{0mm} 8&65  \\
 \hspace{0mm}         \hspace{0mm} &  $^{ 1 }{ \Sigma  }^{ - }$   \hspace{0mm}  & \hspace{0mm} 9&42 \hspace{0mm} & \hspace{0mm} 7&22 \hspace{0mm} & \hspace{0mm} 9&45 \hspace{0mm} & \hspace{0mm} 9&22 \hspace{0mm}  \\
 \hspace{0mm}         \hspace{0mm} &  RMSRD$_4$   \hspace{0mm} & \hspace{0mm} 0&07 \hspace{0mm} & \hspace{0mm} 0&23 \hspace{0mm} & \hspace{0mm} 0&09 \hspace{0mm} & \hspace{0mm} 0&00 \hspace{0mm} \\
\hline
 \hspace{0mm} C$_2$H$_2$  \hspace{0mm} &  $^{ 3 }{ \Sigma  }_{ u }^{ - }$   \hspace{0mm}  & \hspace{0mm} 5&74 \hspace{0mm} & \hspace{0mm} 4&60 \hspace{0mm} & \hspace{0mm} 5&71 \hspace{0mm} & \hspace{0mm} 5&72 \hspace{0mm}  \\
 \hspace{0mm}         \hspace{0mm} &  $^{ 3 }{ \Sigma  }_{ u }^{ - }$   \hspace{0mm}  & \hspace{0mm} 7&28 \hspace{0mm} & \hspace{0mm} 5&73 \hspace{0mm} & \hspace{0mm} 7&28 \hspace{0mm} & \hspace{0mm} 7&16 \hspace{0mm}  \\
 \hspace{0mm}         \hspace{0mm} &  $^{ 1 }{ \Sigma  }_{ u }^{ - }$   \hspace{0mm}  & \hspace{0mm} 8&37 \hspace{0mm} & \hspace{0mm} 6&61 \hspace{0mm} & \hspace{0mm} 8&47 \hspace{0mm} & \hspace{0mm} 8&30 \hspace{0mm}  \\
 \hspace{0mm}         \hspace{0mm} &  $^{ 1 }{ \Delta  }_{ u }$   \hspace{0mm}  & \hspace{0mm} 8&70 \hspace{0mm} & \hspace{0mm} 7&08 \hspace{0mm} & \hspace{0mm} 8&81 \hspace{0mm} & \hspace{0mm} 8&64 \hspace{0mm}  \\
 \hspace{0mm}         \hspace{0mm} &  RMSRD$_4$   \hspace{0mm} & \hspace{0mm} 0&03 \hspace{0mm} & \hspace{0mm} 0&21 \hspace{0mm} & \hspace{0mm} 0&07 \hspace{0mm} & \hspace{0mm} 0&00 \hspace{0mm} \\
\hline
 \hspace{0mm} CH$_2$O  \hspace{0mm} &  $^{ 3 }{ B }_{ 2 }$   \hspace{0mm}  & \hspace{0mm} 3&90 \hspace{0mm} & \hspace{0mm} 3&51 \hspace{0mm} & \hspace{0mm}  3&56 \hspace{0mm} & \hspace{0mm} 3&60 \hspace{0mm}  \\
 \hspace{0mm}         \hspace{0mm} &  $^{ 1 }{ B }_{ 2 }$   \hspace{0mm}  & \hspace{0mm} 4&55 \hspace{0mm} & \hspace{0mm} 4&32 \hspace{0mm} & \hspace{0mm}  3&96 \hspace{0mm} & \hspace{0mm} 3&95 \hspace{0mm}  \\
 \hspace{0mm}         \hspace{0mm} &  $^{ 3 }{ A }_{ 1 } $   \hspace{0mm}  & \hspace{0mm} 6&17 \hspace{0mm} & \hspace{0mm} 4&55 \hspace{0mm} & \hspace{0mm}  6&05 \hspace{0mm} & \hspace{0mm} 6&10 \hspace{0mm}  \\
 \hspace{0mm}         \hspace{0mm} &  $^{ 3 }{ A }_{ 2 } $   \hspace{0mm}  & \hspace{0mm} 9&18 \hspace{0mm} & \hspace{0mm} 9&75 \hspace{0mm} & \hspace{0mm}  8&54 \hspace{0mm} & \hspace{0mm} 8&42 \hspace{0mm}  \\
 \hspace{0mm}         \hspace{0mm} &  $^{ 1 }{ B }_{ 1 } $   \hspace{0mm}  & \hspace{0mm} 10&47 \hspace{0mm} & \hspace{0mm} 9&48 \hspace{0mm} & \hspace{0mm}  9&38 \hspace{0mm} & \hspace{0mm} 9&24 \hspace{0mm}  \\
 \hspace{0mm}         \hspace{0mm} &  RMSRD$_5$   \hspace{0mm} & \hspace{0mm} 0&40 \hspace{0mm} & \hspace{0mm} 0&93 \hspace{0mm} & \hspace{0mm} 0&08 \hspace{0mm} & \hspace{0mm} 0&00 \hspace{0mm} \\
\hline\hline
\end{tabular}
\end{table}






To further test the performance of EOM-JAGP, we performed calculations of vertical excitation energies and RMSRD values
for a number of diatomic (He$_2$, Li$_2$, N$_2$ and CO) and polyatomic (HCN, CH$_2$O, and C$_2$H$_2$) systems,
with results shown in Figure \ref{fig:He2} and Tables \ref{tab:diatomic} and \ref{tab:multiatomic}. 
As one might expect, EOM-JAGP delivers nearly exact results for He$_2$ and Li$_2$ as these consist of very weakly interacting pairs of electrons,
an ideal situation for a pairing theory.
In CH$_2$, where we report excitation energies relative to the lowest singlet, we see both EOM-JAGP's large ground state bias
(large enough that it fails to predict a triplet ground state)
and its difficulty in handling double excitations (the error is larger than 1eV for the ${}^1$A$_1$ (HOMO)$^2\rightarrow\hspace{1mm}$(LUMO)$^2$ excitation).
The latter failure should be put into context, however, as EOM-CCSD has essentially the same difficulty.
CH$_2$'s double excitation is thus a good example of how EOM methods' quality typically degrades as an excited state becomes more and more
different from the ground state, as occurs when increasing numbers of electrons are excited.


In N$_2$, CO, HCN, C$_2$H$_2$ and CH$_2$O, EOM-JAGP displays less of a ground state bias.
As we have not in this study optimized the orbital basis for the Jastrow factor \cite{Neuscamman:2013:or_jagp}, the ground state JAGP is hampered in its
recovery of dynamic correlation, which appears to manifest more strongly in these multiply-bonded systems in which the correlation between electron pairs
is expected to be more important.
We expect that the resulting raising of JAGP's ground state energy is responsible for the observed reduction in ground state bias.
In terms of relative energies (as measured by RMSRD), EOM-JAGP's performance in these molecules is intermediate between CIS and EOM-CCSD,
as one would expect from a simple examination of the size of the methods' derivative subspaces.

\section{Conclusions}
\label{sec:conclusions}

We have presented an equation of motion (EOM) linear response method compatible with variational Monte Carlo.
We find that any ansatz compatible with the ground state linear method is likely to be compatible with this approach.
As an initial example, we pair the formalism with the Jastrow antisymmetric geminal power (JAGP) ansatz, whose EOM subspace
we find to be intermediate in flexibility between that of configuration interaction singles (CIS) and coupled cluster
singles and doubles (CCSD).
Somewhat counterintuitively, the unusually compact (compared to its high accuracy) nature of the JAGP for ground states
leads to a general overestimation of excited state energies.
Nonetheless, we find that in terms of relative energies between excited states, EOM-JAGP is as expected intermediate
in accuracy between CIS and EOM-CCSD in single-reference systems, while performing much more reliably than CIS in
the more multi-reference setting of the carbon dimer and displaying a similarly poor accuracy as EOM-CCSD for
double excitations.
While Monte Carlo's high prefactor makes it difficult to recommend EOM-JAGP for general use in its current form,
the fact that variational Monte Carlo admits both linear response and variational frameworks for excited states
makes it an unusually rich area in which to look for more powerful excited state methods in the future.

\section{Acknowledgments}
\label{sec:acknowledgments}
The authors acknowledge funding from the Office of Science, Office of Basic Energy Sciences,
the US Department of Energy, Contract No. DE-AC02-05CH11231.
Calculations were performed using the Berkeley Research Computing Savio cluster.


\bibliographystyle{aip}
\bibliography{eom_paper.bib}

\end{document}